\documentclass[10pt]{iopart}
\usepackage[]{graphicx}
\usepackage{bm}
\usepackage{enumerate}
\usepackage{balance}
\usepackage{hyperref}
\usepackage{framed}
\usepackage{dcolumn}% Align table columns on decimal point
\usepackage{iopams}
\usepackage{bm}% bold math
\ioptwocol
\begin{document}
\noindent
\JPCM (Accepted 11/2019)\\
\letter{Experimental determination of the bare energy gap of GaAs without the zero-point renormalization}
\author{Basabendra Roy, Kingshuk Mukhuti, Bhavtosh Bansal}
\ead{bhavtosh@iiserkol.ac.in}
\address{Indian Institute of Science Education and Research Kolkata, Mohanpur, Nadia 741246, West Bengal, India}
\begin{abstract}
The energy gap of simple band insulators like GaAs is a strong function of temperature due to the electron-phonon interactions. Interestingly, the perturbation from zero-point phonons is also predicted to cause significant (a few percent) renormalization of the energy gap at absolute zero temperature but its value has been difficult to estimate both theoretically and, of course, experimentally. Given the experimental evidence [Bhattacharya, et al., Phys. Rev. Lett. 114, 047402 (2015)] that strongly supports that the exponential broadening (Urbach tail) of the excitonic absorption edge at low temperatures is the manifestation of this zero temperature electron-phonon scattering, we argue that the location of the Urbach focus is the zero temperature unrenormalized gap. Experiments on GaAs yield the zero temperature bare energy gap to be 1.581 eV and thus the renormalization is estimated to be 66 meV.
\end{abstract}
\section*{Introduction}
In theories like quantum electrodynamics, which describe fundamental interactions, the free fields are an idealization around which renormalized  perturbation theory is implemented to extract physically measurable quantities like the electron mass. An electron stripped of its `photon cloud' has infinite mass and infinite charge and is thus an abstract entity which does not exist \cite{Milonni}.

Electrons in a solid state environment experience similar polaronic dressing on account of electron-phonon interactions \cite{Allen, Cardona, Giustino}. But the presence of a short distance cut-off and the fact that a physical electron can exist outside the crystal environment result in the perturbative corrections being finite. It is thus meaningful to talk about the bare mass and energy values for the electrons in a (hypothetical) static lattice environment. While the effects of the dynamic lattice are most obviously manifest in the temperature dependence of the critical point energies \cite{Giustino, Allen-Nery, bhosale, Lautenschlager, Cardona_PRL2004}, the magnitude of the correction due to the zero-point motion of atoms is also thought to be significant \cite{Giustino, Cannuccia, Thewalt, McKenzie-Wilkins}.

In this work, we have attempted to experimentally determine this zero temperature renormalization of the energy gap in an archetypal band insulator, bulk GaAs.  In the past decade, as more precise theoretical tools have become available, there has been a renewed interest in calculating this zero-point renormalization (ZPR) \cite{Giustino, Cannuccia, Giustino_PRL,Antonius, Gonze,  Monserrat_theory,  Zacharias1,  Ortenzi, Mishra, Friedrich, Ponce_JCP,  Nery-Allen, Nery-Allen2, Tutchton, Karsai}. It is of course impossible to divest the atoms forming the crystal of their zero-point motion and one must employ some stratagem to experimentally determine the unrenormalized gap. While there have been some previous attempts \cite{Thewalt, Monserrat_expt}, which we will describe below, the fundamentally difficult nature of this enterprise has resulted in large uncertainties. Here we demonstrate how the low temperature Urbach edge \cite{Urbach,  Rupak, Sadigh, Soukoulis, Halperin, Cody, Johnson, Greeff, dow-redfield} of a high purity band insulator may be used to directly estimate the bare {\em excitonic} gap \cite{Rupak}.

\begin{center}
\begin{figure}[!b]
\includegraphics[scale=0.40]{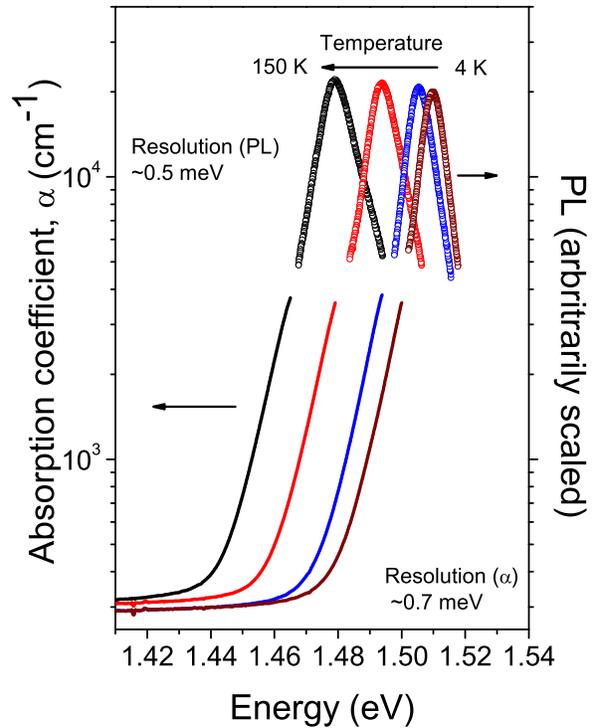}
\caption{Absorption and photoluminescence (PL) measurements from bulk GaAs at 4 K, 50 K, 100 K, and 150 K. Due to the relatively thick sample, the absorption coefficient could not be measured up to the saturation value \cite{Sturge} of $\approx 2-3\times 10^{4}$ cm$^{-1}$. The exponential tail is also evident in the PL spectra. The PL spectra shown in the figure are arbitrarily scaled such that exponential low energy region extrapolates the absorption curves, as expected  \cite{Rupak_APL}.  This sample, labeled $S2$, has $n$-type doping density of $5\times10^{15}$ cm$^{-3}$. While visually inviting, the excitonic band edge cannot be identified with the peak of the PL spectra as there is 5 meV Stokes' shift from the excitonic energy of 1.515 eV.}
\end{figure}
\end{center}
\section*{The Urbach edge}
It has long been known that the joint density of states of almost all band insulators has a `tail' just below the fundamental energy gap. In this bandtail region, the absorption coefficient $\alpha(\hbar\omega, T)$ has a universal form  given by the Urbach rule \cite{Urbach, Rupak}
\begin{equation}\label{Eq:def urbach}
\alpha(\hbar\omega,T)=\alpha _{0}\exp\left[{-\sigma(T)(\varepsilon_{_F}-\hbar \omega)/k_{B}T}\right].
\end{equation}
Here $\alpha(\hbar\omega,T)$ is the absorption coefficient at the incident photon energy $\hbar\omega$ and temperature $T$.  $\varepsilon_{_F}$ and $\alpha_0$, if they are temperature independent constants, define the coordinates of the {\em Urbach focus} \cite{Cody} where the bandtail absorption curves at different temperatures would all extrapolate to meet. $\sigma(T)$, the `Urbach parameter', characterizes the steepness of the absorption edge. Experiments over the past six decades have provided evidence of the universality of the Urbach rule \cite{Urbach, Rupak, Cody, Johnson, Greeff, dow-redfield}, even if the physical origins for the exponential tail within the theoretically forbidden gap may be different \cite{dow-redfield, cohen, toyozawa,Schafer}. For excitonic band edges observed in materials like GaAs, $\sigma(T)$ has a temperature dependence of the form \cite{Schafer, Rupak}
\begin{equation}\label{Eq:DefSigma}
\sigma(T)=\sigma_{0}\frac{2k_{B}T}{\hbar\Omega_{p}}\tanh\frac{\hbar\Omega_{p}}{2k_{B}T}
\end{equation}
where $\sigma_{0}$ is a material dependent constant and $\hbar\Omega_{p}$ represents the energy of phonons involved in the formation of the absorption tail.

We next show that for conventional band insulators like GaAs, the Urbach focus energy $\varepsilon_{_F}$ is (i) temperature independent and (ii) the unrenormalized gap at zero temperature \cite{Cody}. In general, there seems to be no apparent reason for $\varepsilon_{_F}$ in Eq. (\ref{Eq:def urbach}) to be temperature-independent. Indeed, as the band tail is defined with respect to the band edge, the exponential slope at different temperatures should laterally shift following the temperature dependent shift of the band edge, i.e., instead of Eq. (\ref{Eq:def urbach}), we expect \footnote{The argument so presented suggests that one should write  $\alpha(\hbar\omega,T)=\tilde{\alpha}_{0}\exp\left(-[\sigma(T)/k_bT](E_0+E_g(T)-\hbar \omega)\right)$, where the reference energy for the Urbach edge at any given temperature is $E_0+E_g(T)$ rather than simply $E_g(T)$, with $E_0$ being another constant. But note that one is dealing with the excitonic bandedge at low temperature \cite{Sturge} and the exciton line is practically a delta-function in absence of phonon effects and other disorder. Thus for Eq. (\ref{Eq:alphaTempdep}) to have the correct limit $E_0\approx 0$.}
\begin{equation}\label{Eq:alphaTempdep}
\alpha(\hbar\omega,T)=\tilde{\alpha}_{0}\exp\left[{-\sigma(T)(E_g(T)-\hbar \omega)/k_BT}\right].
\end{equation}
 Here we have replaced $\varepsilon_{_F}$ with the temperature dependent gap $E_g(T)$ and $\tilde{\alpha}_{0}$ is some constant different from $\alpha_0$.  We now show that under the adiabatic approximation with a single Einstein mode\footnote{Note that the description in terms of a single Einstein mode does not work for materials where more than one phonon branch shows a TO-LO mode splitting \cite{Monserrat_expt}. Furthermore, the adiabatic approximation itself, despite the weak coupling, is known to severely break down for even for GaAs since it is polar \cite{Ponce_JCP}. This argument is therefore somewhat simplistic and not completely rigorous.}, Eq. (\ref{Eq:alphaTempdep}) and Eq. (\ref{Eq:def urbach}) are in fact identical.

From microscopic theory, the temperature dependence of the energy gap resulting from the electron-phonon interactions, at the simplest level, is described by the adiabatic Allen-Heine formula \cite{Giustino}. The formula is derived on further approximating the second-order perturbation theory expression to get the renormalized energy $E_{j\textbf{k}}$ for the $j^\textrm{th}$ band in the following form:
\begin{equation}\label{Eq:Allen-Heine}
E_{j\textbf{k}}=\varepsilon_{j\textbf{k}}+\sum_{\nu, {\bf q}} \Phi (j, \nu, {\bf q}, {\bf k}) \left[2n_{q\nu}+1 \right].
%E_{n\textbf{k}}=\epsilon_{n\textbf{k}}+\sum_{\nu, {\bf q}}\bigg[\sum_{m}\frac{|g_{mn\nu}(\textbf{k},q)|^{2}}{\epsilon_{n\textbf{k}}-\epsilon_{m\textbf{k}+q}}\\+\;g^{DW}_{nn\nu\nu}(\textbf{k,q,-q})\bigg](2n_{q\nu}+1)
\end{equation}
Here ${\bf q}$ and ${\bf k}$ are respectively the phonon and electron wave vectors, $\varepsilon_{j\textbf{k}}$ is the bare energy of the $j^{th}$ band at wave vector ${\bf k}$, and $\Phi(j, \nu, {\bf q}, {\bf k})$ is the e-p interaction matrix element containing the sum of the Debye-Waller and the $2^{nd}$ order perturbation theory contributions from the $\nu^\textrm{th}$ phonon branch. Since the contribution to the energy renormalization from different phonon branches cannot be physically separated, for the purpose of modeling experimental data, $\nu$ different dispersive phonon branches are all lumped together and effectively considered as a single dispersionless Einstein mode of energy $\hbar \Omega_p$ \cite{Vina, Lautenschlager}. Eq. (\ref{Eq:Allen-Heine}) thus simplifies into an expression for the temperature dependence of critical point energies with only three fitting parameters. The renormalized difference between the valence and the conduction band energies, i.e., the energy gap $E_{g}(T)$ can also thus be expected to have a temperature dependence given by \cite{Vina, Lautenschlager}
\begin{equation}\label{Eq:CardonaFormula}
E_{g}(T)=\varepsilon^{0}_{g}-\Delta E^0_g\coth\left[{\hbar\Omega_{p}\over 2k_{B}T}\right].
\end{equation}
$\varepsilon^{0}_{g}$ is the band gap of the hypothetical phonon-free (bare) crystal and the $\coth (\hbar\Omega_{p}/2k_{B}T)$ is just $[1+2n(\Omega_p, T)]$, with $n(\Omega_p, T)$ being the Bose distribution for the Einstein phonons of frequency $\Omega_p$. Thus $\varepsilon^0_{g}-\Delta E^0_g$ is the experimentally measured bandgap at zero temperature and $\Delta E^0_g$  the zero-point renormalization (ZPR). As the energy gap is lowered by phonon interactions, the minus sign is added to Eq. (\ref{Eq:CardonaFormula}) so that $\Delta E^0_g>0$. Plugging Eq. (\ref{Eq:CardonaFormula}) into Eq. (\ref{Eq:alphaTempdep}), Eq. (\ref{Eq:def urbach}) is recovered if we read $\alpha_0\equiv\exp(\Delta E^0_g 2\sigma_0/\hbar\Omega_p)\tilde{\alpha}_0$. Thus the focus $\varepsilon_{_F}$ is temperature-independent and equal to $\varepsilon^{0}_{g}$, the unrenormalized zero-temperature energy gap.

\begin{center}
\begin{figure}
\includegraphics[scale=0.3]{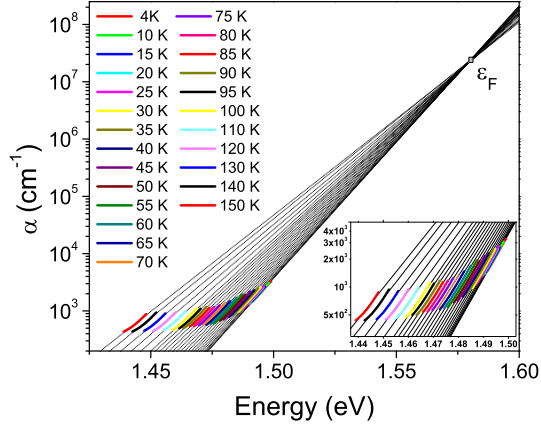}
\caption{Exponential part of the absorption spectrum (Urbach tail) measured from GaAs at different temperatures between 4 K and 150 K  extrapolates to a converging point (Urbach focus). Inset shows the magnified view of the relevant region of the absorption coefficient to highlight the quality of linear fit (on the semilog scale). The data shown here is for the same sample as in figure 1. }
\end{figure}
\end{center}
\section*{Experimental}
Three $n$-type bulk GaAs samples, labeled $S1$, $S2$, and $S3$,  with background carrier densities of $2\times 10^{15}$ cm$^{-3}$, $5\times 10^{15}$ cm$^{-3}$, and $5\times 10^{16}$ cm$^{-3}$ respectively, were chosen for this study. It is conventional to consider samples with $n< 1\times 10^{16}$ cm$^{-3}$ pure enough for determination of fundamental band parameters \cite{Lautenschlager} and three samples with doping varying over an order of magnitude were chosen to rule out any carrier density dependence in this range of doping. The electric field of ionized donors or acceptors affects the Urbach edge only for densities larger than about $5\times 10^{18}$ cm$^{-3}$ \cite{pankove}.

While conceptually straightforward, reliable measurement of the absorption coefficient in a semiconductor is difficult. Due to the four orders of magnitude change in the optical density close to the gap, it is virtually impossible to measure the whole range with a single sample of a given thickness \cite{Sturge}. To get to the value where the excitonic and band to band absorption saturates, i.e., $\alpha>10^4$ cm$^{-1}$, one needs to thin the sample to less than 10 $\mu$m \cite{Sturge}. But such thin samples have their own problems. Being extremely fragile, they need to be stuck on a supporting substrate, making them susceptible to strain in a temperature-dependent measurement. Furthermore, the transmission data is corrupted by Fabry-Perot fringes. Since the Urbach tail occurs in the range $\lesssim 2\times 10^2$ cm$^{-1}\lesssim$ $\alpha \lesssim 2\times 10^3$ cm$^{-1}$, we chemically polished samples thick enough ($\approx 100$ $\mu$m) to be freely supported. A few representative absorption spectra at different temperatures, from sample $S2$ ($n=5\times 10^{15}$ cm$^{-3}$) are shown in figure 1.

To determine the bandtails at different temperatures, a region of the absorption spectrum around $\approx 1000$ cm$^{-1}$ \cite{pankove} was chosen where a straight line could be well-fitted on a semilog scale [figure 2]. The range of the straight line fit was temperature dependent because the absorption edge is corrupted by a low energy shoulder due to the phonon sidebands \cite{Haug-Schmitt-Rink}; the absorption spectrum around the bandedge has additional contributions from acoustic phonon-assisted transitions at higher temperatures.

The Urbach focus energy $\varepsilon_{_F}$ was determined by the following procedure. Firstly, the points of intersection for every pair of such fitted straight lines (on the semilog scale), corresponding to the measured band tails at temperatures, say, $T_i$ and $T_j$ was determined to yield a focus energy $\varepsilon_{_F}^{i,j}$ for this pair. One thus had an ensemble of $^NC_2$ Urbach foci $\lbrace\varepsilon_{_F}\rbrace$ for a given sample, where $N$ is the number of straight lines, each corresponding to a temperature at which the absorption measurement was done. By taking intersection point energies within a window of $10$ meV about the median (which includes more than $90\%$ of the energy values), the histogram for the possible foci (Figure 3) directly yields both the mean as well as the standard deviation which can be used as a rough error estimate.\footnote{The analysis for the other two samples, S1 and S3, is discussed in the Supplementary Data and the results are summarized in Table 1.} The value of the zero temperature excitonic energy gap of GaAs is precisely known to be $1.515$ eV \cite{madelung, meyer_review}. As we see no specific trend in the value of $\varepsilon_{_F}$ with the carrier density in our range of doping ($n<5\times 10^{16}$ cm$^{-3}$), we can combine the estimates made on the three samples to get the value of the unrenormalized gap to be $1.581$ eV and thus the ZPR is about $66$ meV.

How do these values compare with the other estimates? The most obvious way to estimate the ZPR is to directly use Eq. (\ref{Eq:CardonaFormula}) while fitting the temperature dependence of the bandedge. This yields $\Delta E^0_g=57 \pm 29$ meV \cite{Lautenschlager, Zacharias1}. Cardona had also proposed that the unrenormalized gap may be estimated by an extrapolation scheme, based on Eq. (\ref{Eq:CardonaFormula}), that identifies the intercept to a linear fit in the limit of $T\rightarrow\infty$ with $\varepsilon^0_g$. This method yields $54$ meV (with an error that is difficult to estimate) to be the ZPR for GaAs \cite{Thewalt}. An obvious problem with this method has been that one needs to go to the $T\rightarrow\infty$ limit to obtain a $T=0$ result. The fact that the temperature dependence of the bandgap curve is only asymptotically linear implies that the ZPR will be underestimated. A further ambiguity in this method stems from the fact that at higher temperatures, apart from it being no longer excitonic, the absorption would be greatly broadened. Another approximate estimate inferred from the isotope mass shift is $45$ meV \cite{Thewalt}.

Theoretically, the state-of-the-art numerical estimate of ZPR are about $33$ meV \cite{Antonius, Zacharias1} using density functional perturbation theory. However, the use of more refined many-body based approaches \cite{Zacharias1} is expected to yield a theoretical ZPR of about $45$ meV.  The reason for the underestimation of the theoretical values with respect to the present experimental ZPR is not clear at present.

\begin{center}
\begin{figure}
\includegraphics[scale=.3]{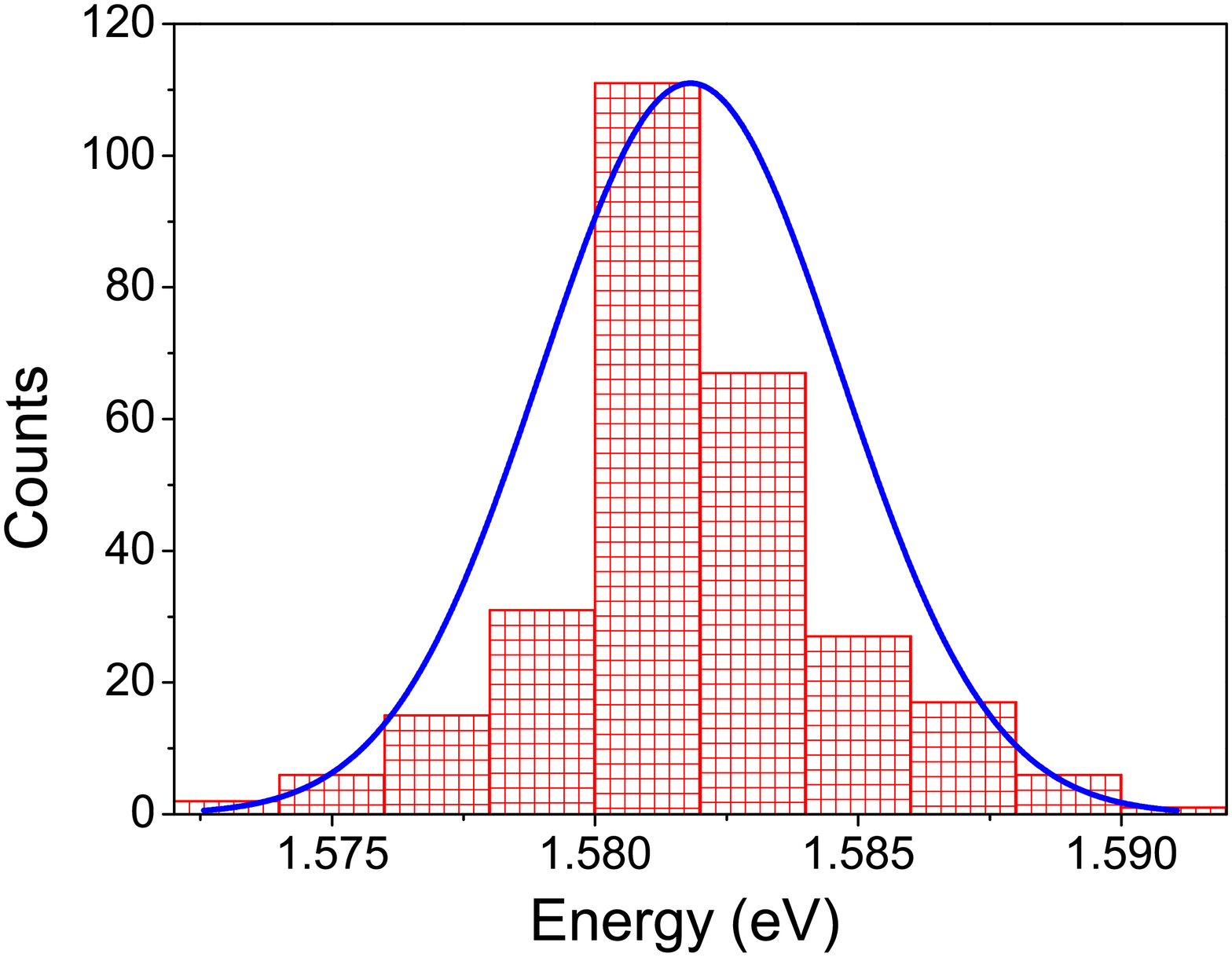}
\caption{Histogram of the values of energy at which different pairs of straight line fits in figure 2 intersect. Fit to the Gaussian function yields the mean value of $\varepsilon_{_F}=1.5884$ eV
with a standard deviation of $0.0034$ eV. This standard deviation is taken to be the error in the location of the focus.}
\end{figure}
\end{center}

\begin{table}
\label{tab:title}
\caption{Parameters extracted from the Urbach edge analysis for three GaAs samples. Here $n$ is the carrier density, $\sigma_0$ is the Urbach slope parameter, $\hbar \Omega_p$ is the inferred optical phonon frequency, $\varepsilon_{_F}$ is the Urbach focus identified with the bare gap, $\Delta E^0_g$ is the ZPR. The numbers in the parenthesis indicate the uncertainty in the last digit.}
  \begin{tabular}{ c   c   c   c   }
    \hline
    \hline
    Sample & $S1$ & $S2$ & $S3$ \\
    \hline
    n (cm$^{-3}$) & $ <2 \times 10^{15}$& $5 \times 10^{15}$& $5 \times 10^{16}$\\
    $\sigma_{0}$ & 1.269 & 1.399  & 1.340 \\
    $\hbar \Omega_p$ (meV) & 25.0 & 25.9 &  26.7\\
   $E_g(0) (eV)$ (PL) & 1.5119(5) & 1.5097(5) & 1.5132 (5)\\
    $\varepsilon_{_F}$ (eV) & $1.579 (3)$ & $1.582(3)$ & $1.582 (4)$\\
    $E_u$ (meV) & 9.84 & 9.26 & 9.95\\
    $\Delta E^0_g$ (meV)$^\dagger$ & $64\pm 3$ & $67\pm 3$ & $67 \pm 4$ \\
    \hline \hline
\end{tabular}

      $^\dagger$Assuming exciton energy $E_g(0)=1.515$ eV at 0 K \cite{meyer_review}.
\end{table}
\section*{Physics of the excitonic Urbach edge}
The universality of the Urbach tail has been a subject for theoretical investigations for over five decades \cite{Soukoulis, Halperin, Cody,dow-redfield,toyozawa, Haug-Schmitt-Rink}, though the discussion has been usually limited to the materials aspects or understanding the universality of the Urbach rule across material systems, especially amorphous semiconductors \cite{Sadigh, Soukoulis, Halperin, Cody}.
In contrast to this, it was only recently shown [using arguments at least as rigorous as those used to derive Eq. (\ref{Eq:Allen-Heine}) and (\ref{Eq:CardonaFormula})] that for materials like GaAs, the band tail at very low temperatures is the direct manifestation of the perturbation caused by the zero-point phonons to the excitonic gap \cite{Rupak}.

The theory is based on the observation that the separation of time scales between the exciton and lattice motion implies that the exciton, at each instant in time, essentially sees a snapshot of the distorted lattice \cite{Schafer}. Due to the electron-phonon coupling, an exciton experiences this distorted lattice as an effective electric field \cite{dow-redfield, Schafer} and the observed Urbach edge is the exciton resonance-enhanced Franz-Keldysh lineshape \cite{Haug-Schmitt-Rink}. Physically, this line broadening can be thought to be caused by the finite ionization probability (lifetime) due to the possibility of tunneling to the energetically degenerate continuum states \cite{dow-redfield, Dow-Redfield_Field}, as is schematically shown in figure 4.  Support to the theory comes from the fact that the experimental estimate in GaAs quantum wells for the zero-point electric field ($\approx 3$ kVcm$^{-1}$) is also in excellent agreement with the theoretical estimate for the Fr\"ohlich interaction \cite{Rupak}. One thus has an excellent basis for Eq. (\ref{Eq:def urbach}) and  (\ref{Eq:DefSigma}).
\begin{center}
\begin{figure}
\includegraphics[scale=.35]{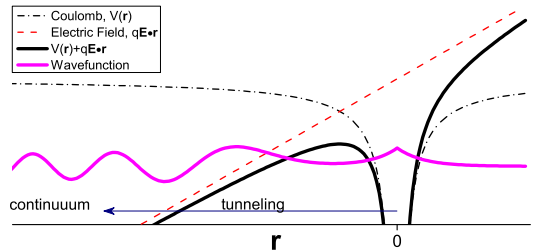}
\caption{Mechanism for Urbach tail when the bandedge is excitonic \cite{Dow-Redfield_Field}. The electron-phonon interaction manifests as a (temperature dependent) electric field that modifies the hydrogenic potential of the bare
exciton by addition of a linear potential.  Since the bound states are now energetically degenerate with the continuum, the additional lifetime broadening of excitonic resonance manifests as the Urbach lineshape \cite{Rupak, dow-redfield, Schafer}.}
\end{figure}
\end{center}

Figure 5 shows the variation of the Urbach slope parameter $\sigma(T)$ with temperature, which when fitted to Eq. (\ref{Eq:DefSigma}) yields the parameters $\sigma_{0}$ and the characteristic phonon energy $\hbar \Omega_{p}$. These are listed in Table I. For comparison, fitting Eq. (\ref{Eq:CardonaFormula}) to the temperature dependent modulated reflectivity was found to yield $27.8$ meV in Ref. \cite{Lautenschlager}. Hence $\Omega_{p}$ in Eq. (2) and Eq. (5) do indeed refer to the same quantity and we have an independent consistency check. An alternate way to characterize the Urbach tail is via the fit to a simple equation $\alpha\sim \exp(\hbar\omega/E_u(T))$, where $E_u(T)$ is the temperature dependent slope. While our measured values of $E_u$ at the lowest temperature are again in excellent agreement with the early data summarized by Pankove where $5<E_u<10 $ meV at helium temperature \cite{pankove}, interestingly, they are about four times larger than what was measured on high quality GaAs quantum wells \cite{Rupak} where $E_u$ was $\approx 2$ meV. This difference, of course, reflects the difference in the excitonic binding energy between bulk and quasi-two dimensional systems.
\begin{center}
\begin{figure}
\includegraphics[scale=.35]{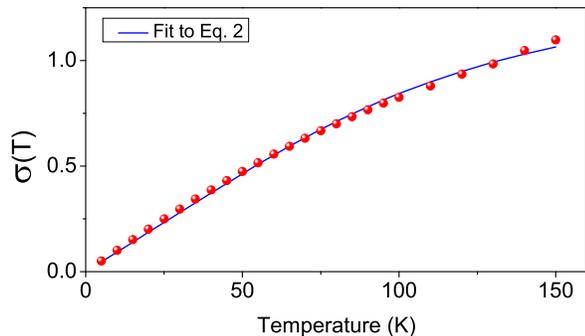}
\caption{Fitting Eq. 2 to the data in figure 2 yields the values of the Urbach slope parameter ($\sigma_0=1.399$) and the characteristic phonon energy ($\hbar\Omega_p=25.9$ meV).}
\end{figure}
\end{center}
\section*{Conclusions}
Simple semiconductors like GaAs are among the best characterized materials we know and effective descriptions like the multiband ${\bf k}\cdot{\bf p}$ theory have been very successful in materials and device engineering. Yet numerically predicting the location of the band edge, quantitative effects of the electron-phonon interactions, and the details of the experimentally measured parameters like the absorption edge continue to be challenging, when attempted {\em ab initio} \cite{bhosale, Antonius, Giustino, Gonze}. In this context, `switching-off' the electron-phonon interaction considerably eases the theory-experiment comparison. The ZPR has thus been a sought after parameter \cite{Giustino,Thewalt,Giustino_PRL, Antonius,Gonze,  Monserrat_theory, Zacharias1, Monserrat_expt} that has so far been difficult to reliably estimate. Given the clean theoretical basis and the simplicity of the measurements, our work shows a way to transcend this difficulty. Similar studies of the Urbach edge can of course yield ZPR in other semiconductors and ZPR should become a fundamental parameter listed in the data tables.

\title[Supplementary Data]{\underline{Supplementary Data}}
\begin{abstract}
We discuss the sample information, details of the absorption and the photoluminescence measurements, and data for the experimental determination of the zero-point renormalization of two other GaAs samples, labeled S1 and S3, with doping densities of $2\times 10^{15}$ cm$^{-3}$  and $5\times 10^{16}$ cm$^{-3}$ respectively.
\end{abstract}
\tableofcontents
\title[Supplementary Data]{}
\section{\label{sec:level1}Sample information and processing}
Three n-type wafers of crystalline GaAs grown by the Czochralski method, labeled S1, S2, and S3 with carrier densities of $1.68\times 10^{15}$ cm$^{-3}$, $5.3\times 10^{15}$ cm$^{-3}$ and $5\times 10^{16}$ cm$^{-3}$ respectively were used in this study. The $\approx 400\,\mu$m thick wafers were chemically etched to a thickness of about $100\,\mu$m such that the resulting pieces also had mirror-like finish on both sides. To thin down the samples we had prepared a solution of H$_{2}$SO$_{4}$ + H$_{2}$O$_{2}$ (30\% strength) + H$_{2}$O in $5:1:1$ ratio [1]. The samples were dipped in the solution allowing a free etch at 330 K which results in successive formation and dissociation of oxides. The mixture is exothermic and the etching rate increases with temperature.
\renewcommand\thefigure{1 (supplement)}
\begin{center}
	\begin{figure}[!b]
		\includegraphics[scale=0.4]{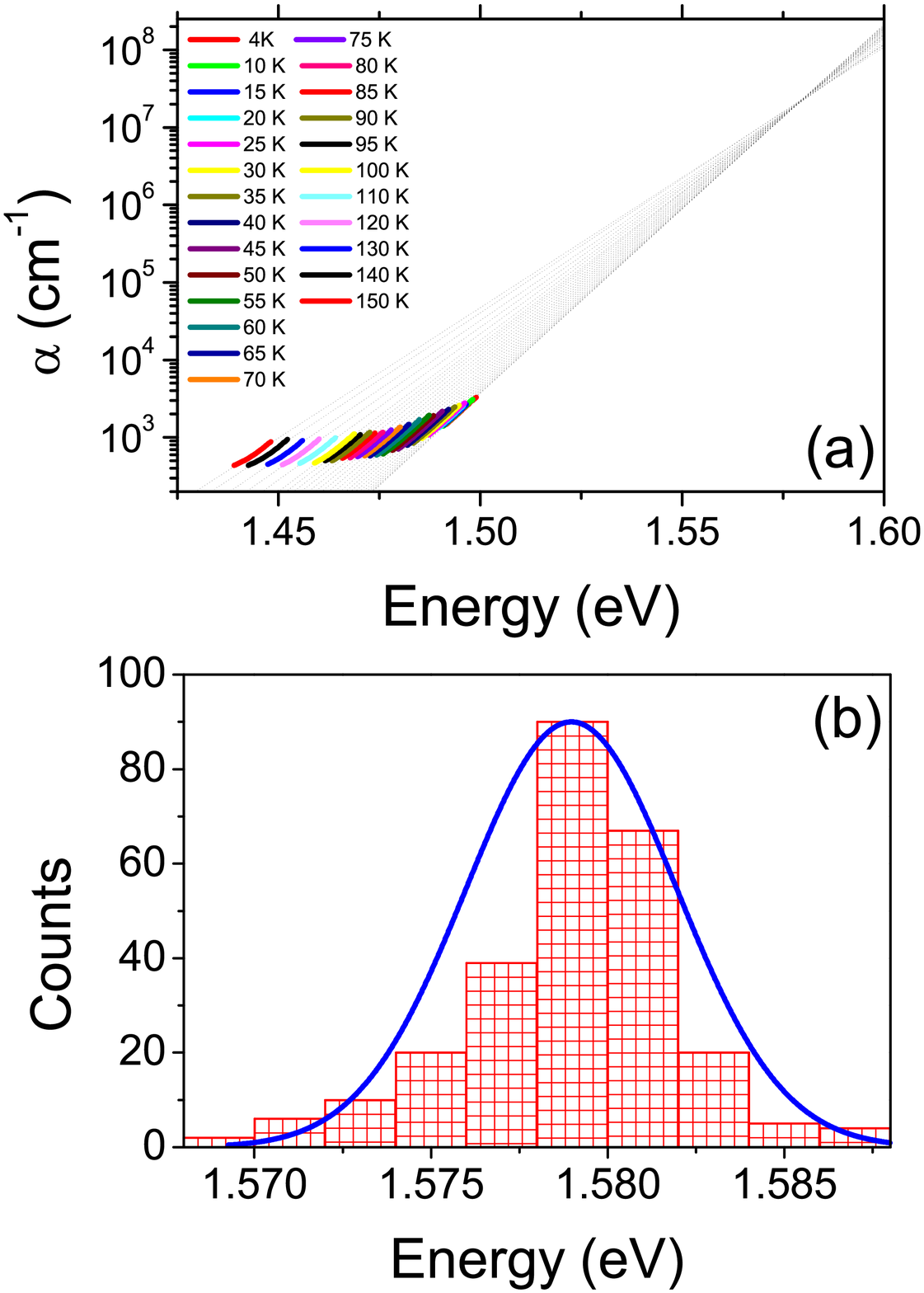}
		\caption{Data on the sample S1 (n=$2\times 10^{15}$cm$^{-3}$)(a) The exponential tails extrapolated to a convergent bundle (b) Histogram of the variation in the Urbach foci. }
	\end{figure}
\end{center}
\renewcommand\thefigure{2 (supplement)}
\begin{center}
	\begin{figure}[!h]
		\centering
		\includegraphics[scale=0.4]{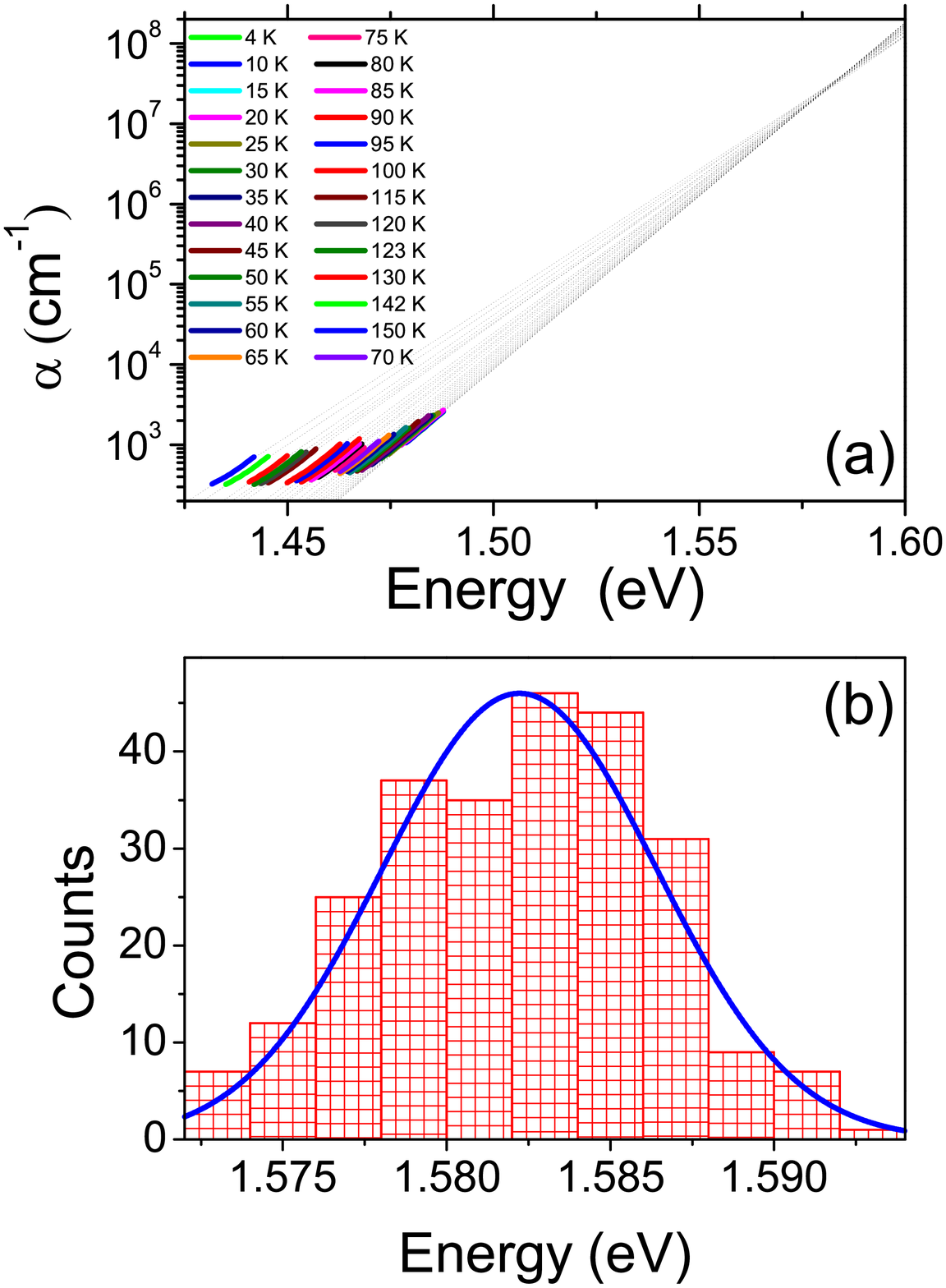}
		\caption{Data on the sample S3 (n=$5\times 10^{16}$cm$^{-3}$) (a) The exponential tails extrapolated to a convergent bundle (b) Histogram of the variation in the Urbach foci }
	\end{figure}
\end{center}
\renewcommand\thefigure{3 (supplement)}
\begin{center}
	\begin{figure}[!h]
		\includegraphics[scale=0.3]{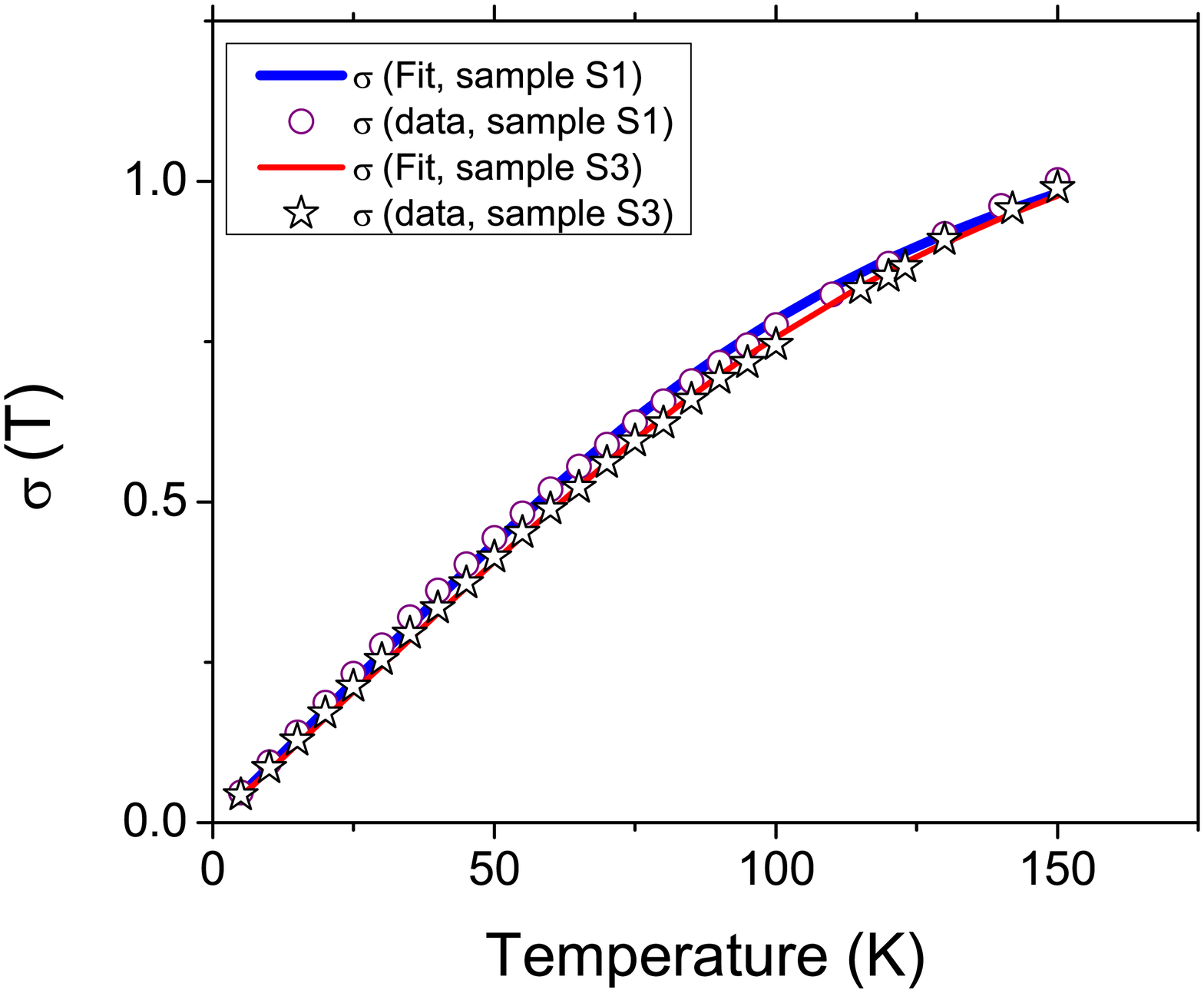}
		\caption{The exponential tails extrapolated to a convergent bundle (b) Histogram of the variation in the Urbach foci  (d) Slope parameter $\sigma$ as a function of temperature, theoretical fit yields a phonon energy of 27.04 meV for sample S1 (n=$2\times 10^{15}$cm$^{-3}$) and 28.72 meV for sample S3 (n=$5\times 10^{16}$cm$^{-3}$).}
	\end{figure}
\end{center}
\section{\label{sec:level2}Absorption measurements}
These etched double-side-polished samples with thickness of about $100\,\mu$m were used in the absorption measurements where the temperature dependent ($4-150$ K) absorption coefficient was determined by transmission measurements using near-infrared radiation from a broadband source (tungsten-halogen lamp) powered by a low noise and stable power supply. The sample was mounted on the cold-finger of a low-vibration helium closed cycle cryostat with optical windows using {\em Apiezon N} grease which is known to have excellent thermal conductivity down to cryogenic temperatures. The spectra were measured using spectrograph connected to an electron-multiplying charge coupled device camera (Andor ixon3) with a spectral resolution of about $0.7$ meV. The spectrograph of calibrated using the sharp lines from the Ne-Ar light source.

For a bulk sample in which the sample thickness is much greater than the wavelength in question and no interference fringes are present, one can write for the absorption spectra taking into account the multiple reflections within the sample [4]

\renewcommand\theequation{a}
 \begin{equation}
\alpha = \frac{1}{d} \log_e\bigg[\frac{2 R^{2} T}{\sqrt{(1-R)^{4} +4 R^{2} T^{2}}-(1-R)^{2}}\bigg]
 \end{equation}
Where $T$ is the total transmittance, $R$ is the reflectance between the sample surface and the surrounding media (air/vacuum), and $d$ is the sample thickness.
Since the reflectivity of the sample has only a very weak wavelength dependence, the reflectivity of each sample was measured at a single wavelength (728 nm) using laser light at close to normal incidence.
\section{Photoluminescence}
Photoluminescence (PL) measurements between $4-150$ K were done using a $728$ nm CW diode laser, with the sample in another helium closed cycle optical cryostat using spectrograph equipped with a charge coupled device camera (Princeton Applied Research Pixis 100). The spectral resolution of the measurement was about $0.5$ meV and the spectrograph was calibrated using the Ne-Ar light source.

The spectral peak position was a non-monotonic function of intensity with the spectra at very low powers and high powers having their peak at lower energy. This is easily understood as being due to the predominance of the defect-bound exciton emission at low powers and heating dependent red-shift of the sample at higher powers. To get the free exciton luminescence, the incident laser power was varied with a variable neutral density filter such that the PL peak was at the highest possible energy. Yet it seems that defect emission always dominates the PL spectrum at $4$ K. All the three samples show a Stokes shift ($2-5$ meV) in the measured PL spectrum with respect to the low temperature excitonic energy ($1.515$ eV) [2, 3]. As expected  [5, 6], the low energy side of the PL spectra is exponential and extrapolate the Urbach edge measured in the absorption spectrum, making the two measurements consistent.

\section{\label{sec:level2}Data and Analysis}
Data on the other two samples, S1 and S3, whose parameters are listed in Table 1 (main text) are shown in Fig. 1 (supplement), 2 (supplement), and 3 (supplement). The analysis is identical to that done for sample S2.
The final unremormalized gap is found to be $1.581\pm 0.002$ eV which yields a zero-point renormalization of $66\pm 2$ meV.
As there is no observed trend in the measured parameters with the doping density (nor is it expected for the present values of doping), the above numbers are estimated by simply taking the mean of the three values of the Urbach foci [see Table 1 (main text)]: $1.579$ eV (for S1), $1.582$ eV (for S2) and $1.582$ eV (for S3). The error is nominally estimated by using the relation $\Delta x={1\over N}\sqrt{\sum_{i=1}^N {\Delta x_i^2}}=2$ meV where $\Delta x_i$ is the uncertainty in the estimate of the $i^{th}$ sample. Note that there may be a slight underestimation in the error because it does not account for the goodness of the fit in Figure 2 (main text) and about 10\% of the data points were discarded while making the histogram.

\section*{References in the Supplementary Data}
\addcontentsline{toc}{section}{\protect\numberline{}References}%
{\small
\begin{enumerate}[1.]
\item S. Sugawara, K. Saito, J. Yamauchi and M. Shoji, Chemical Etching of $\lbrace 111\rbrace$ Surfaces of GaAs Crystals in H$_{2}$SO$_{4}$-H$_{2}$O$_{2}$-H$_{2}$O System, Jap. J. Appl. Phys. {\bf 40}, 12 (2001).
\item I. Vurgaftman, J. R. Meyer, and L. R. Ram-Mohan, Band parameters for III--V compound semiconductors and their alloys,  J.  Appl. Phys. {\bf 89}, 5815 (2001).
\item O. Madelung, Semiconductors: Data Handbook, 3rd ed. p118 (Springer, Berlin 2004).
\item See e.g., S.  R. Johnson  and T. Tiedje, Temperature  dependence  of  the  Urbach  edge  in  GaAs, J. Appl.  Phys. {\bf  78}, 9 (1995).
\item R. Bhattacharya, B. Pal, and B. Bansal, On conversion of luminescence into absorption and the van Roosbroeck-Shockley relation, Appl. Phys. Lett. {\bf 100}, 222103 (2012).
\item  R. Bhattacharya, R. Mondal, P. Khatua, A. Rudra, E. Kapon, S. Malzer, G. D\"ohler, B. Pal, and B. Bansal, Measurements of the Electric Field of Zero-Point Optical Phonons in GaAs Quantum Wells Support the Urbach Rule for Zero-Temperature Lifetime Broadening, Phys. Rev. Lett. {\bf 114}, 04740 (2015).
\end{enumerate}
}
\end{document}